\documentclass{ws-ijmpa}
\usepackage[super,compress]{cite}
\usepackage{graphicx}

\def\Journal#1#2#3#4{{#1} {\bf #2}, #3 (#4)}
\def\APJ{Astrophys. J.}

\def\JHEP{JHEP}

\def\JETPUSSR{JETP (USSR)}
 
\def\MPLA{Mod. Phys. Lett. A}

\def\NPB{Nucl. Phys. B}

\def\NPBSUPPL{Nucl. Phys. B. Proc. Suppl.}

\def\PLB{Phys. Lett. B}

\def\PLBOLD{Phys. Lett.}

\def\PRL{Phys. Rev. Lett.}
\def\PRD{Phys. Rev. D}

\def\PTP{Prog. Theor. Phys.}
\def\RMP{Rev. Mod. Phys.}

\def\SCIENCE{Science}

\def\ZETP{Zh. Eksp. Teor. Piz.}

\begin{document}
\markboth{Teruyuki Kitabayashi and Masaki Yasu\`{e}}{Maximal CP Violation in flavor Neutrino Masses}

\catchline{}{}{}{}{}
%
%%%%%%%%%%%%%%%%%%%%%%%%%%%%%%%%%%%%%%%%%%%%%%%%%%%%%%%%%%%%%%%%%%%%

\title{Maximal CP Violation in Flavor Neutrino Masses}

\author{Teruyuki Kitabayashi and Masaki Yasu\`{e}}

\address{Department of Physics, Tokai University,\\
4-1-1 Kitakaname, Hiratsuka, Kanagawa 259-1292, Japan\\
teruyuki@tokai-u.jp, yasue@keyaki.cc.u-tokai.ac.jp}

\maketitle

\begin{history}
%\received{Day Month Year}
%\revised{Day Month Year}
\end{history}

%%-------------------------------------------------
%% Abstract
%%-------------------------------------------------
\begin{abstract}
Since flavor neutrino masses $M_{\mu\mu,\tau\tau,\mu\tau}$ can be expressed in terms of $M_{ee,e\mu,e\tau}$, mutual dependence among $M_{\mu\mu,\tau\tau,\mu\tau}$ is derived by imposing some constraints on $M_{ee,e\mu,e\tau}$. For appropriately imposed constraints on $M_{ee,e\mu,e\tau}$ giving rise to both maximal CP violation and the maximal atmospheric neutrino mixing, we show various specific textures of neutrino mass matrices including the texture with $M_{\tau\tau}=M^\ast_{\mu\mu}$ derived as the simplest solution to the constraint of $M_{\tau\tau}-M_{\mu\mu}$=imaginary, which is required by the constraint of $M_{e\mu}\cos\theta_{23}-M_{e\tau}\sin\theta_{23}$=real for $\cos 2\theta_{23}=0$. It is found that Majorana CP violation depends on the phase of  $M_{ee}$.

\end{abstract}

\ccode{PACS numbers:14.60.Pq}

%\tableofcontents

%% Main body
%%-------------------------------------------------
%%%%%%%%%%%%%%%%%%%%%%%%%%%%%%%%%%%%%%%%%
\section{\label{sec:1}Introduction}
%%%%%%%%%%%%%%%%%%%%%%%%%%%%%%%%%%%%%%%%%
The observed data of the neutrino oscillations \cite{atmospheric1,atmospheric2,atmospheric3,atmospheric4,solarold1,solarold2,solarold3,solarold4,solarold5,solarold6,solar1,solar2,solar3,solar4,solar5,reactor1,reactor2,reactor3,accelerator1,accelerator2,sin13_1,sin13_2,sin13_3,sin13_4,sin13_5} have been accumulated to suggest the allowed range of the CP-violating Dirac phase $\delta_{CP}$ near $3\pi/2$ inducing the maximal CP violation and that of the atmospheric neutrino mixing angle $\theta_{23}$ near 45$^\circ$ indicating the maximal atmospheric neutrino mixing \cite{NuData1,NuData2}. These maximal effects arouse theoretical interest that both the Dirac CP-violation and the atmospheric neutrino mixing are necessarily maximal because of a certain constraints imposed on flavor neutrino masses. 

It has been discussed that these two maximal effects are correlated \cite{MassTextureCP1_1,MassTextureCP1_2,MassTextureCP1_3,MassTextureCP1_4,MassTextureCP1_5,MassTextureCP1_6,MassTextureCP1_7,MassTextureCP2_1,MassTextureCP2_2,MassTextureCP2_3,MassTextureCP3_1,MassTextureCP3_2,MassTextureCP3_3,MassTextureCP3_4,MassTextureCP3_5,NewMaximalCP_1,NewMaximalCP_2,NewMaximalCP_3,NewMaximalCP_4}.  Denoting the CP-violating Dirac phase by $\delta_{CP}$, the reactor neutrino mixing angle by $\theta_{13}$  and flavor neutrino masses by $M_{ij}$ for $i,j$=$e.\mu,\tau$, we can derive the following relation \cite{MassTextureCP2_1,MassTextureCP2_2,MassTextureCP2_3}:
%%%%%%%%%%%%%%%%%%%%
\begin{equation}
\frac{{{M_{\tau \tau }} - {M_{\mu \mu }}}}{2}\sin 2{\theta _{23}} - {M_{\mu \tau }}\cos 2{\theta_{23}} = \tan {\theta_{13}}\left( {{M_{e\mu }}\cos {\theta_{23}} - {M_{e \tau }}\sin {\theta_{23}}} \right){e^{ - i{\delta _{CP}}}},
\label{Eq:M_nu for 23}
\end{equation}
%%%%%%%%%%%%%%%%%%%%
provided that the Pentecorvo-Maki-Nakagawa-Sakata mixing matrix $U_{PNMS}$ \cite{PMNS1,PMNS2} takes the standard parameterization defined by the Particle Data Group (PDG) \cite{PDG1,PDG2} to be $U_{PDG}=U^{PDG}_\nu K^{PDG}$:
%%%%%%%%%%%%%%%%%%%%
\begin{eqnarray}
U_\nu^{PDG}&=&\left( {\begin{array}{*{20}{c}}
{{c_{12}}{c_{13}}}&{{s_{12}}{c_{13}}}&{{s_{13}}{e^{ - i{\delta _{CP}}}}}\\
{ - {{s_{12}{c_{23}}} - {c_{12}}{s_{23}}{s_{13}}{e^{i{\delta _{CP}}}}} }&{{c_{12}}{c_{23}} - {s_{12}}{s_{23}}{s_{13}}{e^{i{\delta _{CP}}}}}&{{s_{23}}{c_{13}}}\\
{{s_{12}}{s_{23}} - {c_{12}}{c_{23}}{s_{13}}{e^{i{\delta _{CP}}}}}&{ - {c_{12}}{{s_{23}} - {s_{12}}{c_{23}}{s_{13}}{e^{i{\delta _{CP}}}}} }&{{c_{23}}{c_{13}}}
\end{array}} \right),
\nonumber\\
{K^{PDG}} &=& \left( {\begin{array}{*{20}{c}}
{{e^{i{\phi _1}/2}}}&0&0\\
0&{{e^{i{\phi _2}/2}}}&0\\
0&0&{{e^{i{\phi _3}/2}}}
\end{array}} \right),
\label{Eq:U_PDG}
\end{eqnarray}
%%%%%%%%%%%%%%%%%%%%
for $c_{ij}=\cos\theta_{ij}$ and $s_{ij}=\sin\theta_{ij}$ ($i,j$=1,2,3), where $\theta_{12}$ is the solar neutrino mixing angle and $\phi_{1,2,3}$ stand for the Majorana phases, from which two independent combinations become the CP-violating Majorana phases \cite{CPViolationOrg1,CPViolationOrg2,CPViolationOrg3}.  It is obvious that, from Eq.(\ref{Eq:M_nu for 23}), the maximal atmospheric mixing giving $\cos 2\theta_{23}=0$ induces the maximal CP violation giving $\cos \delta_{CP}=0$ if 
%%%%%%%%%%%%%%%%%%%%
\begin{equation}
M_{\tau \tau } - M_{\mu \mu } = {\rm imaginary},
\quad
M_{e\mu}\cos\theta_{23} - M_{e\tau}\sin\theta_{23}\left(=\frac{M_{e\mu}-\sigma M_{e\tau}}{\sqrt{2}} \right)= {\rm real},
\label{Eq:imagenary22-33real12-13}
\end{equation}
%%%%%%%%%%%%%%%%%%%%
where $\sigma=\pm 1$ takes care of the sign of $\sin\theta_{23}$. The reactor neutrino mixing angle must satisfy $\sin {\theta_{13}}\neq 0$.  It is obvious that the simplest solution to satisfy in Eq.(\ref{Eq:imagenary22-33real12-13}) is that 
%%%%%%%%%%%%%%%%%%%%
\begin{equation}
M_{\tau \tau } = M^\ast_{\mu \mu }, \quad
M_{e\tau} = -\sigma M^\ast_{e\mu},
\label{Eq:imagenary22-33real12-13-example}
\end{equation}
%%%%%%%%%%%%%%%%%%%%
which reproduce the following known flavor neutrino mass matrix $M_\nu$ consisting of $M_{ij}$:
%%%%%%%%%%%%%%%%%%%%
\begin{equation}
M_\nu=\left( {\begin{array}{*{20}{c}}
{{M_{ee}}}&{{M_{e\mu }}}&{ - \sigma M_{e\mu }^\ast }\\
{{M_{e\mu }}}&{{M_{\mu \mu }}}&{{M_{\mu \tau }}}\\
{ - \sigma M_{e\mu }^\ast }&{{M_{\mu \tau }}}&{M_{\mu \mu }^\ast }
\end{array}} \right).
\label{Eq:M_nu for maximalCP-23}
\end{equation}
%%%%%%%%%%%%%%%%%%%%

In this article, we would like to discuss the variation of Eq.(\ref{Eq:M_nu for maximalCP-23}) that gives the maximal CP violation and the maximal atmospheric neutrino mixing in the systematic way. In Sec.\ref{sec:2}, we derive $M_{\mu\mu,\tau\tau,\mu\tau}$ expressed in terms of $M_{ee,e\mu,e\tau}$, which are used in Sec.\ref{sec:3} to find how $M_{\mu\mu,\tau\tau,\mu\tau}$ are correlated to each other for $\cos 2\theta_{23}=0$ and $\cos\delta_{CP}=0$. To see the usefulness of our method, we show three new textures other than Eq.(\ref{Eq:M_nu for maximalCP-23}). In the \ref{sec:Appendix}, redundant phases allowed in $U_{PMNS}$ are used to transform $M_\nu$ of Eq.(\ref{Eq:M_nu for maximalCP-23}) into a more general form of $M_\nu$ so that the associated $U_{PMNS}$ \cite{NewMaximalCP_1,NewMaximalCP_2,NewMaximalCP_3,NewMaximalCP_4} is described by three real numbers $u_{1,2,3}$ and three complex numbers $w_{1,2,3}$ \cite{MassTextureCP1_1,MassTextureCP1_2,MassTextureCP1_3,MassTextureCP1_4,MassTextureCP1_5,MassTextureCP1_6,MassTextureCP1_7}. In Sec.\ref{sec:4}, we discuss how these results are affected if $M_\nu$ is not compatible with the PDG convention. The final section Sec.\ref{sec:5} is devoted to summary and discussions.

%%%%%%%%%%%%%%%%%%%%%%%%%%%%%%%%%%%%%%%%%
\section{\label{sec:2}Neutrino Mixings and Flavor Neutrino Masses}
%%%%%%%%%%%%%%%%%%%%%%%%%%%%%%%%%%%%%%%%%
To discuss the situation of the maximal CP violation, we utilize two more relations satisfied by $M_{ij}$ in addition to Eq.(\ref{Eq:M_nu for 23}) \cite{MassTextureCP2_1,MassTextureCP2_2,MassTextureCP2_3}.  The relations determine the remaining two mixing angles, $\theta_{12}$ and $\theta_{13}$.  After little calculus, we obtain that
%%%%%%%%%%%%%%%%%%%%
\begin{eqnarray}
&&
\left( {{\lambda _1} - {\lambda _2}} \right)\sin 2{\theta _{12}} + 2\left( {{c_{23}}{M_{e\mu }} - {s_{23}}{M_{e\tau }}} \right)\frac{{\cos 2{\theta _{12}}}}{{{c_{13}}}} = 0,
\label{Eq:M_nu for 12}\\
&&
\left( {{M_{ee}}{e^{ - i{\delta _{CP}}}} - {\lambda _3}{e^{i{\delta _{CP}}}}} \right)\sin 2{\theta _{13}} + 2\left( {{s_{23}}{M_{e\mu }} + {c_{23}}{M_{e\tau }}} \right)\cos 2{\theta _{13}} = 0,
\label{Eq:M_nu for 13}
\end{eqnarray}
%%%%%%%%%%%%%%%%%%%%
where $\lambda_{1,2,3}$ are defined by
%%%%%%%%%%%%%%%%%%%%
\begin{eqnarray}
{\lambda _1} &=& \frac{{c_{13}^2{M_{ee}} - s_{13}^2{e^{2i{\delta _{CP}}}}{\lambda _3}}}{{\cos 2{\theta _{13}}}},
\nonumber\\
{\lambda _2} &=& c_{23}^2{M_{\mu \mu }} + s_{23}^2{M_{\tau \tau }} - {M_{\mu\tau}}\sin 2{\theta _{23}},
\nonumber\\
{\lambda _3} &=& s_{23}^2{M_{\mu \mu }} + c_{23}^2{M_{\tau \tau }} + {M_{\mu\tau}}\sin 2\theta _{23}.
\label{Eq:lambad-123}
\end{eqnarray}
%%%%%%%%%%%%%%%%%%%%
From these three relations, defining that
%%%%%%%%%%%%%%%%%%%%
\begin{equation}
{M_ + } = {s_{23}}{M_{e\mu }} + {c_{23}}{M_{e\tau }},
\quad
{M_ - } = {c_{23}}{M_{e\mu }} - {s_{23}}{M_{e\tau }},
\label{Eq:MainRelations_MplusMminus}
\end{equation}
%%%%%%%%%%%%%%%%%%%%
we can derive that
%%%%%%%%%%%%%%%%%%%%
\begin{eqnarray}
{M_{\mu \mu }} &=& \left( \frac{1}{c_{13}\tan 2\theta_{12}} - \frac{{{t_{13}}{e^{ - i{\delta _{CP}}}}}}{{\sin 2{\theta_{23}}}} \right)M_- + \left( {\frac{{{e^{ - i{\delta _{CP}}}}}}{{\tan 2{\theta_{13}}}} - \frac{1}{2}{t_{13}}{e^{i{\delta _{CP}}}}} \right)M_+
\nonumber\\
&&
+ \frac{{1 + {e^{ - 2i{\delta _{CP}}}}}}{2}{M_{ee}} - M_{\mu\tau }^{\left( 0 \right)}\cos 2{\theta_{23}},
\nonumber\\
{M_{\tau \tau }} &=& \left( \frac{1}{c_{13}\tan 2\theta_{12}} + \frac{{{t_{13}}{e^{ - i{\delta _{CP}}}}}}{{\sin 2{\theta_{23}}}} \right)M_- + \left( {\frac{{{e^{ - i{\delta _{CP}}}}}}{{\tan 2{\theta_{13}}}} - \frac{1}{2}{t_{13}}{e^{i{\delta _{CP}}}}} \right)M_+
\nonumber\\
&&
+ \frac{{1 + {e^{ - 2i{\delta _{CP}}}}}}{2}{M_{ee}} + M_{\mu\tau }^{\left( 0 \right)}\cos 2{\theta_{23}},
\nonumber\\
{M_{\mu \tau }} &=&  M_{\mu\tau }^{\left( 0 \right)}\sin 2{\theta_{23}},
\label{Eq:MainRelations}
\end{eqnarray}
%%%%%%%%%%%%%%%%%%%%
and
%%%%%%%%%%%%%%%%%%%%
\begin{eqnarray}
M_{\mu\tau }^{\left( 0 \right)} &=& -\left( \frac{1}{{{c_{13}}\tan 2{\theta_{12}}}} + \frac{t_{13}e^{ - i\delta _{CP}}}{\tan 2\theta_{23}} \right)M_- + \left( \frac{{{e^{ - i{\delta _{CP}}}}}}{{\tan 2{\theta_{13}}}} + \frac{1}{2}t_{13}e^{i\delta _{CP}} \right)M_+
\nonumber\\
&&
 - \frac{{1 - {e^{ - 2i{\delta _{CP}}}}}}{2}{M_{ee}}.
\label{Eq:MainRelations_mutau_0}
\end{eqnarray}
%%%%%%%%%%%%%%%%%%%%
where $t_{ij}=\tan\theta_{ij}$ ($i,j=1,2,3)$.  In other words, the derived flavor masses of Eq.(\ref{Eq:MainRelations}) are the solutions to Eqs.(\ref{Eq:M_nu for 23}), (\ref{Eq:M_nu for 12}) and (\ref{Eq:M_nu for 13}).  The relation of 
%%%%%%%%%%%%%%%%%%%%
\begin{equation}
M_+\propto \sin\theta_{13},
\label{Eq:M+_theta_13}
\end{equation}
%%%%%%%%%%%%%%%%%%%%
holds in the limit of $\sin\theta_{13}\rightarrow 0$ as can be seen from Eq.(\ref{Eq:M_nu for 13}).

The neutrino masses $m_{1,2,3}$ are calculated to be:
%%%%%%%%%%%%%%%%%%%%
\begin{eqnarray}
{m_1}{e^{ - i{\phi _1}}} &=&  - \frac{t_{12}}{c_{13}}{M_ - } - t_{13}{e^{i{\delta _{CP}}}}{M_ + } + {M_{ee}},
\nonumber\\
{m_2}{e^{ - i{\phi _2}}} &=& \frac{1}{c_{13}t_{12}}{M_ - } - t_{13}{e^{i{\delta _{CP}}}}{M_ + } + {M_{ee}},
\nonumber\\
{m_3}{e^{ - i{\phi _3}}} &=& \frac{1}{t_{13}}{e^{ - i{\delta _{CP}}}}{M_ + } + {e^{ - 2i{\delta _{CP}}}}{M_{ee}}.
\label{Eq:MainRelationsMass}
\end{eqnarray}
%%%%%%%%%%%%%%%%%%%%
The advantage of our method lies in the fact that the neutrino masses simply depend on the flavor masses of $M_{ee,e\mu,e\tau}$.  It should be noted that somehow ``natural" choice of $M_\pm$ and $M_{ee}$ is to take
%%%%%%%%%%%%%%%%%%%%
\begin{equation}
M_-,M_{ee} = {\rm real},
\quad
\arg\left(M_+\right)= -\delta _{CP},
\label{Eq:MainRelationsMass-1}
\end{equation}
%%%%%%%%%%%%%%%%%%%%
which induce 
%%%%%%%%%%%%%%%%%%%%
\begin{equation}
\phi_{1,2}=0,
\quad
\phi_3=2\delta_{CP}.
\label{Eq:NaturalChoice}
\end{equation}
%%%%%%%%%%%%%%%%%%%%
As a result, the Majorana CP violation is characterized by the phase $2\delta_{CP}$ \cite{ArbitaryCP}.  If textures belong to this type, no Majorana CP violation is induced for the maximal CP violation.

%%%%%%%%%%%%%%%%%%%%%%%%%%%%%%%%%%%%%%%%%
\section{\label{sec:3}Maximal CP Violation and Maximal Atmospheric Neutrino Mixing}
%%%%%%%%%%%%%%%%%%%%%%%%%%%%%%%%%%%%%%%%%
Using $e^{i\delta_{CP}}=\kappa i$ ($\kappa=\pm 1$) for the maximal CP violation and $\cos 2\theta_{23}=0$ for the maximal atmospheric neutrino mixing, the main constraint Eq.(\ref{Eq:M_nu for 23}) turns out to be the simplest one: 
%%%%%%%%%%%%%%%%%%%%
\begin{equation}
{{M_{\tau \tau }} - {M_{\mu \mu }}}= -2\kappa i\tan {\theta_{13}}\left({M_{e\mu }} - \sigma {M_{e\tau }}\right)/{\sqrt 2 }.
\label{Eq:SimplestConstraint}
\end{equation}
%%%%%%%%%%%%%%%%%%%%
The flavor neutrino masses of $M_{\mu\mu,\tau\tau,\mu\tau}$ are given by
%%%%%%%%%%%%%%%%%%%%
\begin{eqnarray}
{M_{\mu \mu }} &=& \left( {\frac{1}{{{c_{13}}\tan 2{\theta_{12}}}} + i\kappa\sigma {t_{13}}} \right)M_- - i\kappa\left( {\frac{1}{{\tan 2{\theta_{13}}}} + \frac{1}{2}{t_{13}}} \right)M_+,
\nonumber\\
{M_{\tau \tau }} &=& \left( {\frac{1}{{{c_{13}}\tan 2{\theta_{12}}}} - i\kappa\sigma {t_{13}}} \right)M_- - i\kappa \left( {\frac{1}{{\tan 2{\theta_{13}}}} + \frac{1}{2}{t_{13}}} \right)M_+,
\nonumber\\
\sigma {M_{\mu \tau }} &=&  - \frac{1}{{{c_{13}}\tan 2{\theta_{12}}}}M_- - i\kappa \left( {\frac{1}{{\tan 2{\theta_{13}}}} - \frac{1}{2}{t_{13}}} \right)M_+ - {M_{ee}},
\label{Eq:TextureCPMaximal}
\end{eqnarray}
%%%%%%%%%%%%%%%%%%%% 
where $M_\pm$ for the maximal atmospheric neutrino mixing are given by 
%%%%%%%%%%%%%%%%%%%%
\begin{equation}
M_+ = \frac{\sigma {{M_{e\mu }} + {M_{e\tau }}}}{{\sqrt 2 }},
\quad
M_- = \frac{{{M_{e\mu }} - \sigma {M_{e\tau }}}}{{\sqrt 2 }}.
\label{Eq:Eq:MainRelations_MplusMminusMaximal}
\end{equation}
%%%%%%%%%%%%%%%%%%%%
We define two real parameters, $x$ and $y$, and one complex parameter, $z$, to be:
%%%%%%%%%%%%%%%%%%%%
\begin{equation}
x = \frac{1}{{\tan 2{\theta _{13}}}} + \frac{1}{2}{t_{13}},
\quad
y = \frac{1}{{\tan 2{\theta _{13}}}} - \frac{1}{2}{t_{13}},
\quad
z = \frac{1}{{{c_{13}}\tan 2{\theta _{12}}}} + i\kappa \sigma {t_{13}},
\label{Eq:MainRelations-xyz}
\end{equation}
%%%%%%%%%%%%%%%%%%%%
leading to
%%%%%%%%%%%%%%%%%%%%
\begin{eqnarray}
{M_{\mu \mu }} &=& z{M_ - } - i\kappa x{M_ + },
\nonumber \\
{M_{\tau \tau }} &=& {z^ * }{M_ - } - i\kappa x{M_ + }
\nonumber \\
\sigma {M_{\mu \tau }} &=&  - \frac{{z + {z^ * }}}{2}{M_ - } - i\kappa y{M_ + } - {M_{ee}}.
\label{Eq:MainRelations_MplusMminusMaximal-xyz}
\end{eqnarray}
%%%%%%%%%%%%%%%%%%%%

It can be found that the texture of Eq.(\ref{Eq:M_nu for maximalCP-23}) is one of the specific textures, where $M_+$ and $M_-$ are appropriately constrained to be $M_+$=imaginary and $M_-=$real.  One may choose any types of constraints on $M_{ij}$ in place of this constraint to find favorite textures.  Some of the examples including Eq.(\ref{Eq:M_nu for maximalCP-23}) are shown as follows:
%%%%%%%%%%%%%%%%%%%%
\begin{enumerate}
%%%%%%%%%
\item For $M_-$=real(=$R_0$ being a real number), 
%%%%%%%%
\begin{equation}
{M_\nu } = \left( {\begin{array}{*{20}{c}}
{{M_{ee}}}&{\frac{{{R_0} + \sigma {M_ + }}}{{\sqrt 2 }}}&{\frac{{ - \sigma \left( {{R_0} - \sigma {M_ + }} \right)}}{{\sqrt 2 }}}\\
{\frac{{{R_0} + \sigma {M_ + }}}{{\sqrt 2 }}}&{z{R_0} - i\kappa x{M_ + }}&{{M_{\mu \tau }}}\\
{\frac{{ - \sigma \left( {{R_0} - \sigma {M_ + }} \right)}}{{\sqrt 2 }}}&{{M_{\mu \tau }}}&{{z^ * }{R_0} - i\kappa x{M_ + }}
\end{array}} \right),
\label{Eq:M_nu for maximalCP-23-1}
\end{equation}
%%%%%%%%%
is obtained and $m_{1,2,3}$ are given by
%%%%%%%%%
\begin{eqnarray}
{m_1}{e^{ - i{\phi _1}}} &=&  - \frac{{{t_{12}}}}{{{c_{13}}}}R{_0} - i\kappa {t_{13}}{M_ + } + {M_{ee}},
\nonumber\\
{m_2}{e^{ - i{\phi _2}}} &=& \frac{1}{{{c_{13}}{t_{12}}}}{R_0} - i\kappa {t_{13}}{M_ + } + {M_{ee}},
\nonumber\\
{m_3}{e^{ - i{\phi _3}}} &=&  - \left( {i\frac{\kappa }{{{t_{13}}}}{M_ + } + {M_{ee}}} \right),
\label{Eq:mass for maximalCP-23-1}
\end{eqnarray}
%%%%%%%%%
where, if $M_+$=imaginary is further imposed, we reach Eq.(\ref{Eq:M_nu for maximalCP-23}) with $M_{e\tau}=-\sigma M^\ast_{e\mu}$ and $M_{\tau\tau}=M^\ast_{\mu\mu}$;
%%%%%%%%%
\item For $M_-$=imaginary(=i$I_0$ with $I_0$ being a real number), 
%%%%%%%%
\begin{equation}
{M_\nu } = \left( {\begin{array}{*{20}{c}}
{{M_{ee}}}&{\frac{{i\left( {{I_0} - i\sigma {M_ + }} \right)}}{{\sqrt 2 }}}&{ - \frac{{i\sigma \left( {{I_0} + i\sigma {M_ + }} \right)}}{{\sqrt 2 }}}\\
{\frac{{i\left( {{I_0} - i\sigma {M_ + }} \right)}}{{\sqrt 2 }}}&{i\left( {z{I_0} - \kappa x{M_ + }} \right)}&{{M_{\mu \tau }}}\\
{ - \frac{{i\sigma \left( {{I_0} + i\sigma {M_ + }} \right)}}{{\sqrt 2 }}}&{{M_{\mu \tau }}}&{i\left( {{z^ * }{I_0} - \kappa x{M_ + }} \right)}
\end{array}} \right),
\label{Eq:M_nu for maximalCP-23-2}
\end{equation}
%%%%%%%%%
is obtained and $m_{1,2,3}$ are given by
%%%%%%%%%
\begin{eqnarray}
{m_1}{e^{ - i{\phi _1}}} &=& - i\left( {\frac{{{t_{12}}}}{{{c_{13}}}}I{_0} + \kappa {t_{13}}{M_ + }} \right) + {M_{ee}},
\nonumber\\
{m_2}{e^{ - i{\phi _2}}} &=& i\left( {\frac{{{t_{12}}}}{{{c_{13}}}}I{_0} - \kappa {t_{13}}{M_ + }} \right) + {M_{ee}},
\nonumber\\
{m_3}{e^{ - i{\phi _3}}} &=&  - \left( {i\frac{\kappa }{{{t_{13}}}}{M_ + } + {M_{ee}}} \right),
\label{Eq:mass for maximalCP-23-2}
\end{eqnarray}
%%%%%%%%%
where, if $M_+$=real is further imposed, we observe that $M_{e\tau}=\sigma M^\ast_{e\mu}$ and $M_{\tau\tau}=-M^\ast_{\mu\mu}$ are satisfied in $M_\nu$:
%%%%%%%%
\begin{equation}
M_\nu=\left( {\begin{array}{*{20}{c}}
{{M_{ee}}}&{{M_{e\mu }}}&{\sigma M_{e\mu }^ * }\\
{{M_{e\mu }}}&{{M_{\mu \mu }}}&{{M_{\mu \tau }}}\\
{\sigma M_{e\mu }^ * }&{{M_{\mu \tau }}}&{ - M_{\mu \mu }^ * }
\end{array}} \right);
\label{Eq:M_nu for maximalCP-23-2-1}
\end{equation}
%%%%%%%%%
\item For $M_+= 0$,
%%%%%%%%
\begin{equation}
M_\nu=\left( \begin{array}{*{20}{c}}
M_{ee}&M_{e\mu}&-\sigma M_{e\mu}\\
M_{e\mu}&\sqrt{2}zM_{e\mu}&-\left(\frac{{z + {z^\ast }}}{\sqrt 2}{M_{e\mu }} +M_{ee}\right)\\
-\sigma M_{e\mu}&-\left(\frac{{z + {z^\ast }}}{\sqrt 2}{M_{e\mu }} +M_{ee}\right)&\sqrt{2}z^\ast M_{e\mu}
\end{array} \right),
\label{Eq:M_nu for maximalCP-23-3}
\end{equation}
%%%%%%%%%
is obtained and $m_{1,2,3}$ are given by
%%%%%%%%%
\begin{eqnarray}
{m_1}{e^{ - i{\phi _1}}} &=&  - \sqrt 2 {t_{12}}\frac{{{M_{e\mu }}}}{{{c_{13}}}} + {M_{ee}},
\nonumber\\
{m_2}{e^{ - i{\phi _2}}} &=& \sqrt 2 \frac{1}{{{t_{12}}}}\frac{{{M_{e\mu }}}}{{{c_{13}}}} + {M_{ee}},
\nonumber\\
{m_3}{e^{ - i{\phi _3}}} &=&  - {M_{ee}},
\label{Eq:mass for maximalCP-23-3}
\end{eqnarray}
%%%%%%%%%
where the mass hierarchy will be the degenerated one realized by $\left|M_{ee}\right|\gg\left|M_{e\mu}\right|$;
%%%%%%%%%
\item For $M_-= 0$, Eq.(\ref{Eq:MainRelationsMass}) leads to $m_1=m_2$, which is not allowed;
%%%%%%%%%
\item For $\sigma M_{\mu\tau} =  -M_{ee}$ giving ${M_ - } = i\kappa \tan 2{\theta _{12}}\left( {c_{13}^2 - 2s_{13}^2} \right){M_ + }/2{s_{13}}$,
%%%%%%%%
\begin{equation}
M_\nu=\left( {\begin{array}{*{20}{c}}
M_{ee}&M_{e\mu }& M_{e\tau}\\
M_{e\mu}&-i\sqrt 2 \kappa t_{13}M_{e\tau}& - \sigma M_{ee}\\
M_{e\tau}& - \sigma M_{ee}& - i\sqrt 2 \kappa\sigma t_{13}M_{e\mu}
\end{array}} \right),
\label{Eq:M_nu for maximalCP-23-4}
\end{equation}
%%%%%%%%%
is obtained and  $m_{1,2,3}$ are given by
%%%%%%%%%
\begin{eqnarray}
{m_1}{e^{ - i{\phi _1}}} &=&  - i\kappa \left( {{t_{12}}\tan 2{\theta_{12}}\frac{{1 - 2t_{13}^2}}{{2{t_{13}}}} + {t_{13}}} \right)M_+ + {M_{ee}},
\nonumber\\
{m_2}{e^{ - i{\phi _2}}} &=& i\kappa \left( {\frac{1}{{{t_{12}}}}\tan 2{\theta_{12}}\frac{{1 - 2t_{13}^2}}{{2{t_{13}}}} - {t_{13}}} \right)M_+ + {M_{ee}},
\nonumber\\
{m_3}{e^{ - i{\phi _3}}} &=&  - \left( i\kappa{\frac{1}{{{t_{13}}}}M_+ + {M_{ee}}} \right),
\label{Eq:mass for maximalCP-23-4}
\end{eqnarray}
%%%%%%%%%
where the mass hierarchy will be also the degenerated one realized by $\left|M_{ee}\right|\gg\left|M_+\right|$.
%%%%%%%%%
\end{enumerate}
%%%%%%%%%%%%%%%%%%%%
There are other textures based on other constraints on $M_\nu$.

For the first texture with $M_{e\tau}=-\sigma M_{e\mu}^\ast$ and $M_{\tau\tau}=M_{\mu\mu}^\ast$, no Majorana CP violation is induced if $M_{ee}$ is real as indicated by Eq.(\ref{Eq:mass for maximalCP-23-1}).  It is noted in the  \ref{sec:Appendix} that $U_{PMNS}$ including redundant phases can be parameterized by three real numbers $u_{1,2,3}$ and three complex numbers $w_{1,2,3}$ numbers \cite{MassTextureCP1_1,MassTextureCP1_2,MassTextureCP1_3,MassTextureCP1_4,MassTextureCP1_5,MassTextureCP1_6,MassTextureCP1_7}:
%%%%%%%%
\begin{equation}
{U_{Maximal}} = \left( {\begin{array}{*{20}{c}}
{{u_1}}&{{u_2}}&{{u_3}}\\
{{w_1}}&{{w_2}}&{{w_3}}\\
{w_1^\ast }&{w_2^\ast }&{w_3^\ast }
\end{array}} \right).
\label{Eq:general UPMNS}
\end{equation}
%%%%%%%%%
To reach $U_{Maximal}$, $M_{ee}$ is constrained to be real and no Majorana CP violation is induced.  It is also noted in the \ref{sec:Appendix} that the second texture with $M_{e\tau}=\sigma M_{e\mu}^\ast$ and $M_{\tau\tau}=-M_{\mu\mu}^\ast$ cannot be obtained from any phase transformation of the first texture with $M_{e\tau}=-\sigma M_{e\mu}^\ast$ and $M_{\tau\tau}=M_{\mu\mu}^\ast$ by adjusting redundant phases present in $U_{PMNS}$. In this texture, no Majorana CP violation is induced if $M_{ee}$ is imaginary as indicated by Eq.(\ref{Eq:mass for maximalCP-23-2}).

%%%%%%%%%%%%%%%%%%%%%%%%%%%%%%%%%%%%%%%%%
\section{\label{sec:4}Flavor Neutrino Masses with Arbitrary Phases}
%%%%%%%%%%%%%%%%%%%%%%%%%%%%%%%%%%%%%%%%%
For a general form of $M_\nu$, $M_{\mu\mu, \mu\tau, \tau\tau}$ does not satisfy Eq.(\ref{Eq:MainRelations}). In fact, in terms of the redundant phases $\rho$, $\gamma$ and $\tau$ introduced in the \ref{sec:Appendix} yielding $\delta_{CP}=\delta+\rho+\tau$, $M_\nu$ should be replaced by Eq.(\ref{Eq:Mnu for PMNS}), where $M_{ij}$ ($i,j=e,\mu,\tau$) for the PDG convention. Conversely, for a given general form of $M_\nu$, $M_\nu$ compatible with the PDG convention should be
%%%%%%%%%%%%%%%%%%%%
\begin{equation}
M^{PDG}_\nu=\left( {\begin{array}{*{20}{c}}
{{e^{2i\rho }}{M_{ee}}}&{{e^{i\left( {\rho  + \gamma } \right)}}{M_{e\mu }}}&{{e^{i\left( {\rho  - \gamma  - \tau } \right)}}{M_{e\tau }}}\\
{{e^{i\left( {\rho  + \gamma } \right)}}{M_{e\mu }}}&{{e^{2i\gamma }}{M_{\mu \mu }}}&{{e^{ - i\tau }}{M_{\mu \tau }}}\\
{{e^{i\left( {\rho  - \gamma  - \tau } \right)}}{M_{e\tau }}}&{{e^{ - i\tau }}{M_{\mu \tau }}}&{{e^{ - 2i\left( {\gamma  + \tau } \right)}}{M_{\tau \tau }}}
\end{array}} \right).
\label{Eq:generalMnu}
\end{equation}
%%%%%%%%%%%%%%%%%%%%
All constraints on $M_\nu$ including the key relation of Eq.(\ref{Eq:M_nu for 23}) should be replaced by those on $M^{PDG}_\nu$.  We can readily factor out $e^{-i\tau}$ as a common factor to have $\delta  + \tau/2$, $\rho  + \tau/2$ and $\gamma  + \tau/2$, respectively, redefined to be $\delta$, $\rho$ and $\gamma$ resulting in $\delta_{CP}=\delta+\rho$.  

The flavor neutrino masses $M_{\mu\mu, \mu\tau, \tau\tau}$ turn out to satisfy
%%%%%%%%%%%%%%%%%%%%
\begin{eqnarray}
{e^{2i\gamma }}{M_{\mu \mu }} &=& \left( {\frac{1}{{{c_{13}}\tan 2{\theta _{12}}}} - \frac{{{t_{13}}{e^{ - i{\delta _{CP}}}}}}{{\sin 2{\theta _{23}}}}} \right){M_ - } + \left( {\frac{{{e^{ - i{\delta _{CP}}}}}}{{\tan 2{\theta _{13}}}} - \frac{1}{2}{t_{13}}{e^{i{\delta _{CP}}}}} \right){M_ + } 
\nonumber\\
&&
+ \frac{{1 + {e^{ - 2i{\delta _{CP}}}}}}{2}{e^{2i\rho }}{M_{ee}} - M_{\mu \tau }^{\left( 0 \right)}\cos 2{\theta _{23}},
\nonumber\\
{e^{ - 2i\gamma }}{M_{\tau \tau }} &=& \left( {\frac{1}{{{c_{13}}\tan 2{\theta _{12}}}} + \frac{{{t_{13}}{e^{ - i{\delta _{CP}}}}}}{{\sin 2{\theta _{23}}}}} \right){M_ - } + \left( {\frac{{{e^{ - i{\delta _{CP}}}}}}{{\tan 2{\theta _{13}}}} - \frac{1}{2}{t_{13}}{e^{i{\delta _{CP}}}}} \right){M_ + }
\nonumber\\
&&
+ \frac{{1 + {e^{ - 2i{\delta _{CP}}}}}}{2}{e^{2i\rho }}{M_{ee}} + M_{\mu \tau }^{\left( 0 \right)}\cos 2{\theta _{23}},
\label{Eq:MainRelationsGeneral}
\end{eqnarray}
%%%%%%%%%%%%%%%%%%%%
with the same equation of ${M_{\mu \tau }} = M_{\mu \tau }^{\left( 0 \right)}\sin 2{\theta _{23}}$, where
%%%%%%%%%%%%%%%%%%%%
\begin{eqnarray}
{M_ + } &=& {s_{23}}{e^{i\left( {\rho  + \gamma } \right)}}{M_{e\mu }} + {c_{23}}{e^{i\left( {\rho  - \gamma } \right)}}{M_{e\tau }},
\quad
{M_ - } = {c_{23}}{e^{i\left( {\rho  + \gamma } \right)}}{M_{e\mu }} - {s_{23}}{e^{i\left( {\rho  - \gamma } \right)}}{M_{e\tau }},
\nonumber\\
M_{\mu \tau }^{\left( 0 \right)} &=&  - \left( {\frac{1}{{{c_{13}}\tan 2{\theta _{12}}}} + \frac{{{t_{13}}{e^{ - i{\delta _{CP}}}}}}{{\tan 2{\theta _{23}}}}} \right){M_ - } + \left( {\frac{{{e^{ - i{\delta _{CP}}}}}}{{\tan 2{\theta _{13}}}} + \frac{1}{2}{t_{13}}{e^{i{\delta _{CP}}}}} \right){M_ + }
\nonumber\\
&&
 - \frac{{1 - {e^{ - 2i{\delta _{CP}}}}}}{2}{e^{2i\rho }}{M_{ee}}.
\label{Eq:MainRelationsGeneral_M+M-mutau_0}
\end{eqnarray}
%%%%%%%%%%%%%%%%%%%%

Practically speaking, $\delta$ , $\rho$  and $\gamma$ can be calculated from ${\bf M}_{ij}$  ($i,j=e,\mu,\tau$) for ${\bf M}=M^\dagger_\nu M_\nu$.  The phases $\rho$ and $\delta$ are given by
%%%%%%%%%%%%%%%%%%%%
\begin{equation}
\rho = \arg(X), 
\quad
\delta = -\arg(Y), 
\label{EqExact-delta-rho}
\end{equation}
%%%%%%%%%%%%%% 
for
%%%%%%%%%%%%%% 
\begin{equation}
X = {e^{i\gamma} c_{23} {\bf M}_{e\mu} -  e^{ - i\gamma } s_{23} {\bf M}_{e\tau}},
\quad
Y = e^{i\gamma} s_{23} {\bf M}_{e\mu} +  e^{ - i\gamma }c_{23} {\bf M}_{e\tau},
\label{Eq:XY}
\end{equation}
%%%%%%%%%%%%%% 
where $\gamma$ can be derived from
%%%%%%%%%%%%%% 
\begin{eqnarray}
&&
\cos 2\gamma {\rm Im} \left( {\bf M}_{\mu\tau} \right) - \sin 2\gamma {\rm Re}\left( {\bf M}_{\mu\tau} \right) = t_{13} \sin\delta_{CP}\left|X\right|,
\label{Eq:Exact-gamma}
\end{eqnarray}
%%%%%%%%%%%%%% 
with $|X|=\Delta m^2_{21}\sin 2\theta_{12}/2$ $(\Delta m^2_{21}=m^2_2-m^2_1)$. More details can be found in Refs.\citen{BabaYasue1,BabaYasue2,BabaYasue3,XingZhou}.

It is observed that the maximal atmospheric neutrino mixing and maximal CP violation are realized by the same constraint on $M_{\tau\tau}$: $M_{\tau\tau} = M_{\mu\mu}^\ast$ and by the modified one on $M_{e\tau}$: 
%%%%%%%%%%%%%% 
\begin{equation}
M_{e\tau} =  - \sigma e^{ - 2i\rho}M_{e\mu}^\ast.  
\label{Eq:etauGeneral}
\end{equation}
%%%%%%%%%%%%%% 
The additional constraint that $M_{ee}$=real is replaced by $e^{2i\rho }M_{ee}$=real.

%%%%%%%%%%%%%%%%%%%%%%%%%%%%%%%%%%%%%%%%%
\section{\label{sec:5}Summary}
%%%%%%%%%%%%%%%%%%%%%%%%%%%%%%%%%%%%%%%%%
We have demonstrated the usefulness of flavor neutrino masses expressed in terms of $M_{ee,e\mu,e\tau}$.  Appropriate constraints on $M_{ee,e\mu,e\tau}$ reveal the necessary mutual dependence among $M_{\mu\mu,\tau\tau,\mu\tau}$ to have the maximal CP violation. For example, we have stressed the significant role of the relation: $\left( {{M_{\tau \tau }} - {M_{\mu \mu }}} \right)\sin 2{\theta _{23}}/2 - {M_{\mu \tau }}\cos 2{\theta _{23}} = \tan {\theta _{13}}\left( {{M_{e\mu }}\cos {\theta _{23}} - {M_{e \tau }}\sin {\theta _{23}}} \right){e^{ - i{\delta _{CP}}}}$.  If we impose a constraint on $M_{e\mu,e\tau}$ such as the constraint of ${M_{e\mu }} -\sigma {M_{e \tau }}$=real for the maximal atmospheric neutrino mixing signaled by $\cos 2\theta_{23}=0$, we observe that the maximal CP violation signaled by $\cos\delta_{CP}=0$ is induced by the constraint of ${M_{\tau \tau }} - {M_{\mu \mu }}$=imaginary. The simplest solution to satisfy both constraints consists of $M_{e\tau}=-\sigma M^\ast_{e\mu}$ and $M_{\tau\tau}=M^\ast_{\mu\mu}$, which provide the known texture \cite{MassTextureCP1_1,MassTextureCP1_2,MassTextureCP1_3,MassTextureCP1_4,MassTextureCP1_5,MassTextureCP1_6,MassTextureCP1_7,MassTextureCP2_1,MassTextureCP2_2,MassTextureCP2_3}. 
%%%%%%%%%%%%%%%%%%NEW START 
If $M_\nu$ is not associated with the PDG convention, $M_{e\tau} =  - \sigma M_{e\mu}^\ast$ should be replaced by $M_{e\tau} =  - \sigma e^{ - 2i\rho}M_{e\mu}^\ast$, where $\rho$ is the redundant Dirac phase associated with the 1-2 rotation.
%%%%%%%%%%%%%%%%%%NEW END
Other constraints on $M_{ij}$ ($i,j=e,\mu,\tau$) lead to new textures and we have shown three such examples. It can also be discussed how $M_\nu$ of Eq.(\ref{Eq:M_nu for maximalCP-23}) is modified by the inclusion of the effect from $\cos 2\theta_{23}\neq 0$ allowing an arbitrary atmospheric mixing angle. For example, at a first glance, we may choose the constraint of ${M_{e\mu }}\cos {\theta _{23}} - {M_{\mu \tau }}\sin {\theta _{23}}$=real leading Eq.(\ref{Eq:M_nu for maximalCP-23}) for the maximal atmospheric neutrino mixing.  This subject will be discussed elsewhere.

\appendix

%%-------------------------------------------------
%% Appendix
%%-------------------------------------------------
\section{\label{sec:Appendix}$U_{PMNS}$ for the maximal CP violation and the maximal atmospheric neutrino mixing}

The general form of $U_{PMNS}$ contains seven phases: three phases of the Dirac type to be denoted by $\delta$, $\rho$ and $\tau$, three phases of the Majorana type to be denoted by $\varphi_{1,2,3}$ and the remaining phase to be denoted by $\gamma$ \cite{BabaYasue1,BabaYasue2}.  For $U_{PMNS} = U_\nu K$, we parameterize $U_\nu$ and $K$ as follows:
%%%%%%%%%%%%%%%%%%%%
\begin{eqnarray}
{U_\nu } &=& \left( {\begin{array}{*{20}{c}}
1&0&0\\
0&{{e^{i\gamma }}}&0\\
0&0&{{e^{ - i\gamma }}}
\end{array}} \right)\left( {\begin{array}{*{20}{c}}
1&0&0\\
0&{\cos {\theta _{23}}}&{\sin {\theta _{23}}{e^{i\tau }}}\\
0&{ - \sin {\theta _{23}}{e^{ - i\tau }}}&{\cos {\theta _{23}}}
\end{array}} \right)\left( {\begin{array}{*{20}{c}}
{\cos {\theta _{13}}}&0&{\sin {\theta _{13}}{e^{ - i\delta }}}\\
0&1&0\\
{ - \sin {\theta _{13}}{e^{i\delta }}}&0&{\cos {\theta _{13}}}
\end{array}} \right)
\nonumber\\
&&
\cdot \left( {\begin{array}{*{20}{c}}
{\cos {\theta _{12}}}&{\sin {\theta _{12}}{e^{i\rho }}}&0\\
{ - \sin {\theta _{12}}{e^{ - i\rho }}}&{\cos {\theta _{12}}}&0\\
0&0&1
\end{array}} \right),
\nonumber\\
K &=& \left( {\begin{array}{*{20}{c}}
{{e^{i{\varphi _1}/2}}}&0&0\\
0&{{e^{i{\varphi _2}/2}}}&0\\
0&0&{{e^{i{\varphi _3}/2}}}
\end{array}} \right).
\label{Eq:UPMNS}
\end{eqnarray}
%%%%%%%%%%%%%%%%%%%%
This unitary matrix can be casted into
%%%%%%%%%%%%%%%%%%%%
\begin{eqnarray}
U_{PMNS}=\left( {\begin{array}{*{20}{c}}
{{e^{i\rho }}}&0&0\\
0&{{e^{i\gamma }}}&0\\
0&0&{{e^{ - i\left( {\gamma  + \tau } \right)}}}
\end{array}} \right){U_{PDG}},
\label{Eq:UPMNSvsUPD}
\end{eqnarray}
%%%%%%%%%%%%%%%%%%%%
where
%%%%%%%%%%%%%%%%%%%%
\begin{eqnarray}
{\delta _{CP}} = \delta  + \rho  + \tau,
\quad
{\phi _1} = {\varphi _1} - 2 \rho,
\quad
{\phi _2} = {\varphi _2},
\quad
{\phi _3} = {\varphi _3} + 2 \tau,
\label{Eq:PDGphase}
\end{eqnarray}
%%%%%%%%%%%%%%%%%%%%
in $U_{PDG}$.  Accordingly, from $M_\nu$ for $U_{PDG}$ consisting of $M_{ij}$ ($i,j$=$e,\mu,\tau$), we reach $M_\nu$ for $U_{PMNS}$ of Eq.(\ref{Eq:UPMNS}):
%%%%%%%%%%%%%%%%%%%%
\begin{equation}
{M_\nu } = \left( {\begin{array}{*{20}{c}}
{{e^{ - 2i\rho }}{M_{ee}}}&{{e^{ - i\left( {\rho  + \gamma } \right)}}{M_{e\mu }}}&{{e^{ - i\left( {\rho  - \gamma  - \tau } \right)}}{M_{e\tau }}}\\
{{e^{ - i\left( {\rho  + \gamma } \right)}}{M_{e\mu }}}&{{e^{ - 2i\gamma }}{M_{\mu \mu }}}&{{e^{i\tau }}M_{\mu \tau }}\\
{{e^{ - i\left( {\rho  - \gamma  - \tau } \right)}}{M_{e\tau }}}&{{e^{i\tau }}{M_{\mu \tau }}}&{{e^{2i\left( {\gamma  + \tau } \right)}}{M_{\tau \tau }}}
\end{array}} \right).
\label{Eq:Mnu for PMNS}
\end{equation}
%%%%%%%%%%%%%%%%%%%%

For the maximal CP violation and the maximal atmospheric neutrino mixing, Eq.(\ref{Eq:M_nu for maximalCP-23}) provides
%%%%%%%%%%%%%%%%%%%%
\begin{equation}
{M_\nu } = \left( {\begin{array}{*{20}{c}}
{{e^{ - 2i\rho }}{M_{ee}}}&{{e^{ - i\left( {\rho  + \gamma } \right)}}{M_{e\mu }}}&{ - \sigma {e^{ - i\left( {\rho  - \gamma  - \tau } \right)}}M_{e\mu }^ * }\\
{{e^{ - i\left( {\rho  + \gamma } \right)}}{M_{e\mu }}}&{{e^{ - 2i\gamma }}{M_{\mu \mu }}}&{{e^{i\tau }}{M_{\mu \tau }}}\\
{ - \sigma {e^{ - i\left( {\rho  - \gamma  - \tau } \right)}}M_{e\mu }^ * }&{{e^{i\tau }}{M_{\mu \tau }}}&{{e^{2i\left( {\gamma  + \tau } \right)}}M_{\mu \mu }^ * }
\end{array}} \right).
\label{Eq:MaximalMnu for PMNS}
\end{equation}
%%%%%%%%%%%%%%%%%%%%
Since $\rho$ and $\tau$ are redundant, we choose $\rho$ and $\tau$ to satisfy
${e^{ i\left( {2\rho  + \tau } \right)}} = \eta$ ($\eta =\pm 1$) yielding \cite{NewMaximalCP_1,NewMaximalCP_2,NewMaximalCP_3,NewMaximalCP_4}
%%%%%%%%%%%%%%%%%%%%
\begin{equation}
{M^\prime_\nu } = {e^{ - 2i\rho }}\left( {\begin{array}{*{20}{c}}
{{M_{ee}}}&{{e^{i\left( {\rho  - \gamma } \right)}}{M_{e\mu }}}&{ - \eta \sigma {{\left( {{e^{i\left( {\rho  - \gamma } \right)}}{M_{e\mu }}} \right)}^ \ast }}\\
{{e^{i\left( {\rho  - \gamma } \right)}}{M_{e\mu }}}&{{e^{2i\left( {\rho  - \gamma } \right)}}{M_{\mu \mu }}}&{\eta {M_{\mu \tau }}}\\
{ - \eta \sigma {{\left( {{e^{i\left( {\rho  - \gamma } \right)}}{M_{e\mu }}} \right)}^ \ast }}&{\eta {M_{\mu \tau }}}&{{{\left( {{e^{2i\left( {\rho  - \gamma } \right)}}{M_{\mu \mu }}} \right)}^ \ast }}
\end{array}} \right),
\label{Eq:MaximalMnu for PMNS-2}
\end{equation}
%%%%%%%%%%%%%%%%%%%%
The corresponding $U_{PMNS}$ is simply given by
%%%%%%%%%%%%%%%%%%%%
\begin{equation}
{U_{PMNS}} = e^{i\rho }\left( {\begin{array}{*{20}{c}}
{{u_1}{e^{i{\phi _1}/2}}}&{{u_2}{e^{i{\phi _2}/2}}}&{ - i\kappa {u_3}{e^{i{\phi _3}/2}}}\\
{{w_1}{e^{i{\phi _1}/2}}}&{{w_2}{e^{i{\phi _2}/2}}}&{ - i\kappa \sigma {w_3}{e^{i{\phi _3}/2}}}\\
{ - \eta \sigma w_1^ * {e^{i{\phi _1}/2}}}&{ - \eta \sigma w_2^ * {e^{i{\phi _2}/2}}}&{i\kappa \eta w_3^ * {e^{i{\phi _3}/2}}}
\end{array}} \right),
\label{Eq:NewPMNS for MaximalCP}
\end{equation}
%%%%%%%%%%%%%%%%%%%%
where $u_{1,2,3}$ and $w_{1,2,3}$ are given by
%%%%%%%%%%%%%%%%%%%%
\begin{eqnarray}
&&
{u_1} = {c_{12}}{c_{13}},
\quad
{u_2} = {s_{12}}{c_{13}},
\quad
{u_3} = {s_{13}},
\nonumber\\
&&
{w_1} =  - {e^{i\left( {\gamma  - \rho } \right)}}\frac{{{s_{12}} + i\kappa \sigma {c_{12}}{s_{13}}}}{{\sqrt 2 }},
\quad
{w_2} = {e^{i\left( {\gamma  - \rho } \right)}}\frac{{{c_{12}} - i\kappa \sigma {s_{12}}{s_{13}}}}{{\sqrt 2 }},
\nonumber \\
&&
{w_3} = {e^{i\left( {\gamma  - \rho } \right)}}\frac{{i\kappa {c_{13}}}}{{\sqrt 2 }}.
\label{Eq:u123w123}
\end{eqnarray}
%%%%%%%%%%%%%%%%%%%%
The Majorana phases $\phi_{1,2,3}$ are so determined to satisfy Eq.(\ref{Eq:MainRelationsMass}).  The global phase $\rho$ in Eqs.(\ref{Eq:MaximalMnu for PMNS-2}) and (\ref{Eq:NewPMNS for MaximalCP}) are cancelled each other in $U_{PMNS}^T{M^\prime_\nu }{U_{PMNS}}$. 

To reach $U_{Maximal}$ in Eq.(\ref{Eq:general UPMNS}), it is further required that Eq.(\ref{Eq:NaturalChoice}) be satisfied; namely, $\phi_1=\phi_2 = 0$ and $\phi _3 = 2\delta_{CP}(=\pm \pi)$. This choice subsequently requires that $M_{ee}=$real (as indicated by Eq.(\ref{Eq:MainRelationsMass})), which is nothing but the additional constraint imposed in Refs.\citen{MassTextureCP1_1,MassTextureCP1_2,MassTextureCP1_3,MassTextureCP1_4,MassTextureCP1_5,MassTextureCP1_6,MassTextureCP1_7}. The resulting $U_{PMNS}$ of Eq.(\ref{Eq:UPMNSvsUPD}) turns out to be:
%%%%%%%%%%%%%%%%%%%%
\begin{equation}
U_{PMNS}=
{e^{i\rho }}\left( {\begin{array}{*{20}{c}}
{{u_1}}&{{u_2}}&{{u_3}}\\
{{w_1}}&{{w_2}}&{\sigma {w_3}}\\
{ - \eta \sigma w_1^ \ast }&{ - \eta \sigma w_2^ \ast }&{ - \eta w_3^ \ast }
\end{array}} \right).
\label{Eq:PMNS for MaximalCP}
\end{equation}
%%%%%%%%%%%%%%%%%%%%
The obtained $U_{PMNS}$ is literally coincident with $U_{Maximal}$ of Eq.(\ref{Eq:general UPMNS}) for $\sigma=1$ and $\eta=-1$.  It is thus proved that the texture for $U_{Maximal}$ can be derived from the texture for $U_{PDG}$. No Majorana CP violation is induced because of $\phi_1=\phi_2 = 0$ and $\phi _3=\pm \pi$ \cite{NewMaximalCP_3}. 

One may wonder if the texture of Eq.(\ref{Eq:M_nu for maximalCP-23-2}) with $M_{e\tau}=\sigma M_{e\mu}^\ast$ and $M_{\tau\tau}=-M_{\mu\mu}^\ast$ is obtained from Eq.(\ref{Eq:MaximalMnu for PMNS}) for the texture with $M_{e\tau}=-\sigma M_{e\mu}^\ast$ and $M_{\tau\tau}=M_{\mu\mu}^\ast$ by adjusting the redundant phases. Requiring that $M^\prime_{\tau\tau}=-M^{\prime\ast}_{\mu\mu}$ realized by ${e^{i\left( {2\rho  + \tau } \right)}} = i\eta$, we reach
%%%%%%%%%%%%%%%%%%%%
\begin{equation}
{M^\prime_\nu } = {e^{ - 2i\rho }}\left( {\begin{array}{*{20}{c}}
{{M_{ee}}}&{{e^{i\left( {\rho  - \gamma } \right)}}{M_{e\mu }}}&{ - i\eta \sigma {{\left( {{e^{i\left( {\rho  - \gamma } \right)}}{M_{e\mu }}} \right)}^ \ast }}\\
{{e^{i\left( {\rho  - \gamma } \right)}}{M_{e\mu }}}&{{e^{2i\left( {\rho  - \gamma } \right)}}{M_{\mu \mu }}}&{i\eta {M_{\mu \tau }}}\\
{ - i\eta \sigma {{\left( {{e^{i\left( {\rho  - \gamma } \right)}}{M_{e\mu }}} \right)}^ \ast }}&{i\eta {M_{\mu \tau }}}&{ - {{\left( {{e^{2i\left( {\rho  - \gamma } \right)}}M_{\mu \mu }} \right)}^ \ast }}
\end{array}} \right),
\label{Eq:MaximalMnu for PMNS-3}
\end{equation}
%%%%%%%%%%%%%%%%%%%%
indicating $M^\prime_{e\tau}=-i\eta\sigma M^{\prime\ast}_{e\tau}$, which cannot be equivalent to $M_{e\tau}=\sigma M_{e\mu}^\ast$.  The texture with $M_{e\tau}=\sigma M_{e\mu}^\ast$ and $M_{\tau\tau}=-M_{\mu\mu}^\ast$ is not connected to Eq.(\ref{Eq:MaximalMnu for PMNS}).

%%--------------------------------
%%References
%%--------------------------------
%\begin{thebibliography}{000} %for 3 digits
%\begin{thebibliography}{00}  %for 2 digits

\end{document}